# Zeno Freezing and Anti-Zeno Acceleration of the Dynamic Evolution of Acoustic Topological Boundary States

Xiao-Meng Zhang[1], Ze-Guo Chen[1,2*], Guancong Ma[3*], Ming-Hui Lu[1,2,4*] and Yan-Feng Chen[1,2,4]

[1]School of Materials Science and Intelligent Engineering, Nanjing University, Suzhou 215163, China

[2]National Laboratory of Solid State Microstructures, Nanjing University, Nanjing 210093, China

[3]Department of Physics, Hong Kong Baptist University, Kowloon Tong, Hong Kong, China

[4]College of Engineering and Applied Sciences, Nanjing University, Nanjing 210093, China

* Authors to whom correspondence should be addressed: zeguoc@nju.edu.cn; phgcma@hkbu.edu.hk; luminghui@nju.edu.cn

## *Abstract*

Quantum measurements severely disrupt the dynamic evolution of a quantum system by collapsing the probabilistic wavefunction. This principle can be leveraged to control quantum states by effectively freezing the system's dynamics or enhancing transitions between states. These are known as the quantum Zeno effect (ZE) and anti-Zeno effect (AZE), respectively. However, it remains elusive how quantum measurements affect topological states, which are famous for their robustness against disorder and perturbations. Here, we theoretically and experimentally show that the dynamic evolution of topological boundary states (TBSs) can be controlled by quantum-like measurement (QLM). Our work is based on spatially modulated topological acoustic waveguide systems with varying parameters that adiabatically pump the TBS across the bulk to the opposite boundary. Therein, the QLM is emulated using a perturbation to the Hamiltonian known as the Zeno subspace. With the help of quantum metrics, we identify the general conditions for ZE and AZE, and experimentally demonstrate their effects in freezing and accelerating the tunneling of the TBS. Furthermore, we discover a tunneling mechanism by varying the strength of the QLM. These results highlight QLM as a versatile tool for manipulating topological states and wave propagation.

Classical measurements, if executed carefully, do not affect the system dynamics. In comparison, quantum measurements drastically alter system dynamics by collapsing the probabilistic wavefunction [1]. Such an effect can be leveraged to manipulate quantum states. Through repeated measurements, the evolution of a quantum state can be suppressed or accelerated. These phenomena are known as the quantum Zeno effect (ZE) [2–4] and the quantum anti-Zeno effect (AZE) [5,6], respectively. As such, quantum measurements are not only tied to information acquisition but are also useful for a wide range of functionalities, such as enhancing quantum gate fidelity [7–9], teleportation [10,11], error correction [12,13] and incoherent qubit control [14–17]. The effects of quantum measurements can be projected onto a multidimensional subspace of the Hamiltonian, denoted "Zeno subspace". Through this approach, ZE is harnessed for great flexibility in quantum state manipulation [18–21]. This approach also enables classical systems to mimic quantum ZE through Hamiltonian engineering, thereby vastly expanding the application potential of Zeno dynamics in manipulating classical waves. For example, previous studies have simulated state collapse through structural modulation to mimic quantum-like measurements and achieve Zeno dynamics [22–25]. It should be noted that these works do not concern the aspect of information acquisition in measurements.

The Zeno subspace is a restricted region of the Hilbert space where frequent measurements confine the system's state, effectively freezing its evolution. On a different front, topological matters are known for their robustness against changes to system parameters. In particular, topologically protected edge states have gained widespread attention for their robustness against a wide range of perturbations such as disorder and defects. The effect of quantum measurements on the evolution of topological states is, therefore, an intriguing open question. It is also unclear whether the anti-Zeno effect occurs that accelerates the decay of the topological state, and if so, does it decay to bulk states or another topological state?

In this study, we theoretically and experimentally investigate the effect of quantum-like measurement (QLM) on the evolution of topological states realized in acoustic waveguide systems emulating Schrödinger-like dynamics. By introducing repeated perturbations of different configurations to the waveguide systems, we successfully observe Zeno freezing and Zeno acceleration of the evolution of TBS. These effects emulate wavefunction collapse in QLMs by projecting the states to different Zeno subspaces. We further propose a mechanism for generating topological state pumping through strength-varied measurements. Our results demonstrate a novel scheme for wave manipulation via Zeno dynamics and offer insights for rapidly controlling topological states.

*Schemes to realize Zeno freezing.* We first illustrate the principle of QZE and its emulation in acoustics. Fig. 1(a) shows the effect of repeated quantum measurements on the two-level system with Rabi oscillation, which is the oscillation between states $|0\rangle$ and $|1\rangle$ with a period $T$. Under Rabi oscillation, the probability of measuring the system in state $|0\rangle$ at time $t$ is $P = \cos^2(\pi t/T)$. We introduce $n$ evenly spaced ideal instantaneous measurements in the time interval $t = 0.5T$. The final population is a function of $n$: $P(n) = \cos^{2n}(\pi/2n) \approx e^{-\pi^2/4n}$ [3,4] when $n$ it is sufficiently large. This result means that repeatedly measuring blocks the evolution by freezing the system at the initial state.

To extend this concept for classical wave systems, the key is to mimic a wavefunction collapse induced by a QLM. We start with two coupled waveguides (labeled by WG1 and WG2) representing a two-level system. Under parallel-axis approximation, sound waves in the system can be described by the Schrödinger-type equation:

$$-i\,\partial_z |\psi(z)\rangle = \mathcal{H}(z)|\psi(z)\rangle, \tag{1}$$

where $\mathcal{H}(z) = \begin{pmatrix} \beta & \kappa \\ \kappa & \beta \end{pmatrix}$ is the Hamiltonian and $\beta$ is the propagation constant of the first-order guiding mode under a working frequency, and $\kappa$ denotes the coupling between the two modes [26]. The initial state is prepared as $|\psi(0)\rangle = (1,0)^{\mathrm{T}}$. Repeated QLM is introduced to WG2 as a perturbation occurring at a short distance interval $L_m$. Here we set $L_m = 0.02L$ with $L$ being half of the Rabi period length. This is modeled as

$$\mathcal{M}_1 = \begin{pmatrix} 0 & 0 \\ 0 & \Delta\beta \end{pmatrix}, \tag{2}$$

where $\Delta\beta$ denotes the QLM strength. Equation (2) is achieved by attaching segmented waveguides on WG2. The perturbation is introduced repeatedly with proportion $\alpha = nL_m/L$, i.e., $n$ QLMs are performed over the distance $L$. The effect of QLMs is easily illustrated by calculating the population as a function of measurement strength $\Delta\beta$ and measurement times N. The results in Fig. 1(b) imply that frequent strong measurements will induce the ZE effect. It is intuitively seen by projecting the evolution trajectories on the Bloch sphere, as shown in Fig. 1(c). These trajectories illustrate the transition from free evolution with Rabi oscillations to Zeno freezing via frequent and strong measurement that repeatedly collapse the wavefunction to the initial state. This effect can be realized in an acoustic system shown in Fig. 1(d). The system is two rectangular waveguides coupled by an air channel [26]. The perturbation induced by QLM is implemented by attaching segmented sections to WG2. Fig. 1(e) presents simulation and experimental results confirming that continuous QLM induces frozen transport and confines the state within WG1. Thus, QLM can serve as a tool to implement binary operations by controlling the output state. We can utilize Zeno dynamics to restrict the system to evolve along a particular Zeno subspace and achieve flexible control of acoustic waves, enabling them to propagate, split, and even form patterns along predefined paths, as elaborated in the Supplemental Materials [27].

*Zeno Freezing of Acoustic Topological Boundary States.* Next, we study how QLMs can modulate the dynamic transfer of topological boundary state (TBS) driven by synthetic parameters [28–34], a mechanism akin to the well-known Thouless pump [35]. We examine an Aubry-André-Harper model described by the Hamiltonian $\mathcal{H}_0(\varphi) = \sum_m^M (\beta c_m^\dagger c_m + \kappa_{m,m+1} c_{m+1}^\dagger c_m + \kappa_{m+1,m} c_m^\dagger c_{m+1})$, where $c_m^\dagger(c_m)$ is creation (annihilation) operator and $\kappa_{m,m+1} = \kappa_{m+1,m}$ denotes the hopping. The system is a Chern insulator with a quantized magnetic flux appearing as a cosine modulation to the hopping: $\kappa_{m,m+1}(\varphi) = \kappa_0 + \kappa_m \cos(2\pi\ell m + \varphi)$, where $\kappa_0$ and $\kappa_m$ are the static coupling effect and modulation strength, $\ell = \frac{1}{3}$ is the spatial modulating frequency, $\varphi$ is modulation phase that serves as a synthetic dimension.

The model can be realized in an acoustic system consisting of $M = 9$ waveguides coupled by air channels, as shown in Fig. 2(a). Detailed description of this system can be found in ref. [27]. By adjusting the positions of the air channels, we can precisely tune the inter-waveguide interactions to achieve the desired modulations. Two TBSs protected by a non-zero Chern number can be found localized at the two edges of the waveguide array [28], and their propagations along the waveguide ($z$-direction) can be controlled by varying $\varphi$ as a function of $z$. To show the Zeno freezing effect, we choose a profile of $\varphi(z) = \varphi_0 + \delta\varphi \frac{z}{L}$ with $\varphi_0 = 0.25$ as starting point and $\delta\varphi = -0.5$ as the modulation interval shown in the band structure in Fig. 2(b). In the absence of any measurement, this profile adiabatically pumps an initial TBS state localized at the right edge, denoted as $|\psi(0)\rangle = |\psi_R\rangle$, to tunnel across the bulk lattice to the opposite side over a distance of $L$, resulting in an end state $|\psi(L)\rangle = |\psi_L\rangle$ at the left edge. We further impose continuous QLM to modulate the tunneling process and the corresponding total Hamiltonian is $\mathcal{H}(\varphi(z), \Delta\beta(z)) = \mathcal{H}_0(\varphi(z)) + \mathcal{M}(\Delta\beta(z))$, where $\mathcal{M} = c_m^\dagger c_m \Delta\beta(z)$ denotes the continuous QLM applied on the $m$th site whose strength is experimentally tunable by adjusting the attached waveguide height $w$. Here we set $m = 1$. The tunneling effect can be captured through the quantum metric tensor $g = \text{Re}(\langle \partial\psi_j | \partial\psi_j \rangle - \langle \partial\psi_j | \psi_j \rangle \langle \psi_j | \partial\psi_j \rangle)$, which quantifies the "distance" between neighboring eigenstates belonging to the same $j$-th band in the parameter space spanned by synthetic dimension and measurement strength [36–38]. Typically, the value of $g$ is influenced by the gap size between this band and other bands. Its distribution is shown in Fig. 2(b) as a function of $w$ and $\varphi$ [39]. We first follow the path indicated by the red arrow at $w = 0$, corresponding to the case with no measurement. The path crosses a region where $g$ peaks, which indicates that the eigenstate undergoes a drastic change. Indeed, driven by the change of $\varphi$ along this path, the TBS tunnels from the right boundary to the left [28,30]. Next, we introduce the QLM by setting $w = 3.5$ mm. The path is moved north to a region where $g$ is nearly constant in $\varphi$, indicating that the TBS remain unchanged. In other words, the QLM suppresses the evolution that pumps the TBS to the opposite edge so it remains localized at the same boundary. This is the Zeno freezing effect.

To experimentally observe the Zeno freezing, we fabricate the two waveguide configurations, as shown in Fig. 2(c). We set the sample length to be 1 m to satisfy the adiabatic condition verified by applying the Landau-Zener formula [27,30]. The corresponding full-wave simulations and experimental results are presented in Fig. 2(d). Without measurement, the tunneling of TBS is clearly observed. In the presence of continuous QLMs, the evolution is blocked and the TBS remains at the right boundary.

*Anti-Zeno Acceleration of Acoustic Topological Boundary States.* The presence of additional eigenstates in the bulk bands allows measurements to establish a Zeno subspace that conversely accelerate the state decay rather than merely freezing it. This is the called the AZE, as illustrated in Fig. 3(a) using our topological waveguide system. To investigate the condition for such anti-Zeno effect to appear, we study how initial state affected by both QLM and synthetic dimension driven evolution. We prepare the initial state at the input end, $|\Psi(z=0)\rangle = |\psi_s\rangle$, the wavefunction at distance $z$ can be written as [5]

$$|\Psi(z)\rangle = \Omega_s(z)e^{-ik_s z}|\psi_s\rangle + \sum_i \Omega_i(z)e^{-ik_i z}|\psi_i\rangle \qquad (3)$$

$|\psi_s\rangle$ and $|\psi_i\rangle$ are orthonormal eigenvectors of $\mathcal{H}_0(\varphi(0))$ with eigenvalues $k_s$ and $k_i$. This indicates the wavefunction at any distance $z$ is a superposition of the state $|\psi_s\rangle$ and other states $|\psi_i\rangle$, including both bulk states and the TBS at the opposite boundary. The coefficients $\Omega_s(z)$ and $\Omega_i(z)$ represent the weights of $|\psi_s\rangle$ and $|\psi_i\rangle$, respectively, as functions of the propagation distance $z$.

To obtain $\Omega_s(z)$ to gauge the effect of QLM, we apply the adiabatic condition and track the evolution of $\Omega_s(z)$ by projecting the system's eigenstate at position $z$ onto the initial state. We define a relative decay rate $\chi$ to capture the influence of the continuous measurement $\Delta\beta(z)$,

$$\chi(\varphi(z), \Delta\beta(z)) = -\log\left(\frac{|\langle\psi_j(0)\,|\,\Psi_j(z)\rangle|^2}{|\langle\psi_j(0)\,|\,\psi_j(z)\rangle|^2}\right) \qquad (4)$$

where $|\psi_j^0\rangle, |\psi_j^z\rangle$ and $|\Psi_j^z\rangle$ are the $j$-th eigenvectors of $\mathcal{H}_0(\varphi(0))$, $\mathcal{H}_0(\varphi(z))$ and $\mathcal{H}(\varphi(z), \Delta\beta(z))$, respectively. $\chi \ll 0$ ($\chi \gg 0$) corresponds to the ZE (AZE) region, where the state remains "frozen" (undergoes rapid decay) in its initial configuration due to frequent external perturbations, thus preserving (or diminishing) coherence and fidelity.

We plot a phase diagram with $\varphi_0, \delta\varphi$ and $\Delta\beta$ as coordinates to calculate the relative decay rate [40]. This allows us to evaluate how adding QLM $\Delta\beta$ affects the decay rate of the initial state in system with given $(\varphi_0, \delta\varphi)$. Here we focus on $|\psi_s\rangle = |\psi_R\rangle$ as the representative initial state and discussions of other cases are included in ref. [27]. The phase diagram delineating the two distinct phases is presented in Fig. 3(b), with the ZE region consistent with prior findings. We further investigate the AZE region in the phase diagram by choosing $(\varphi_0, \delta\varphi) = (-0.6, 0.4)$. This evolution path is shown in Fig. 3(c), wherein, in the absence of QLM, it lies in a region for nearly constant $g$. When QLM is in presence, the path can traverse a peak in $g$, indicating the QLM induces significant change in the state's composition during the evolution. To further show that $|\psi_R\rangle$ is transferred to the opposite TBS $|\psi_L\rangle$ instead of other bulk states, we introduce an indicator $\zeta$ to represent state localization, defined as normalized intensity difference between left and right unit cells $\zeta = \sum_m^{left}|\psi_j(m)|^2 - \sum_m^{rig}|\psi_j(m)|^2$. As $\zeta \to 1(-1)$, the state is strongly localized at the left (right) boundary. The band structure evolution as well as the $\zeta$ distribution of the eigenstate confirm that $|\psi_R\rangle$ is indeed transferred to the boundary state $|\psi_L\rangle$, as shown in Fig. 3(c). We design another two samples with ($w = 3.5$ mm) and without QLM applied to experimentally verify the AZE. As demonstrated by both simulations and experiments, continuous QLM accelerates state tunneling, resulting in near-perfect population transfer, as shown in Fig. 3(d).

*Topological state tunneling induced by dynamic measurement.* Another key observation from the phase diagram involves a pronounced state transfer occurring at $\delta\varphi = 0$ by varying only the measurement strength, as illustrated in Fig. 4(a). In previous studies, TBS tunneling arises from the variation of $\varphi$ that requires continuous modulation of every waveguide in the entire array [29–33]. In contrast, here we fix $\varphi$ and trigger TBS tunneling by relying purely on AZE with increasing measurement strength. The effect of such a maneuver can be seen again in the quantum metric distribution in Fig.

4(b). The TBS follows the vertical arrow and traverses a path crossing a maximum in $g$. Compared to the approach that pumps $\varphi$ (the horizontal arrow in Fig. 4(b)), although the end parameters are different, the end states are on the same side of the $g$-ridge, an indication that TBS tunneling has occurred for both. By treating $w$ as an additional system parameter, the QLM-induced tunnel can alternatively be understood via state evolution along a gapped, tilted 1D Dirac point shown in Fig. 4(c) and analyzed using the Landau-Zener formula [27].

Compared to the "conventional" tunneling approach, this approach is superior in that it only requires the modulation of a single waveguide to realize state transfer, which is considerably easier to achieve experimentally. We demonstrated this scenario by exciting the boundary state ($\varphi = -0.2$). By gradually varying the measurement strength, i.e., slowly increasing the width $w$ of the first waveguide from 0 to 3.5 mm over a distance of 0.7 m, we experimentally confirmed complete energy transfer to the opposite boundary, as shown in Fig. 4(d). Detailed results of the QLM-induced tunneling effect for different initial states are provided in the Supplementary Materials. Nevertheless, our findings demonstrate successful population transfer driven by dynamically modulating measurement strength.

*Conclusion* We introduce QLM as a degree of freedom to control the evolution of acoustic topological states. By designing perturbative structures, we simulate QLM that induces ZE and AZE that freezes and accelerates the pumping of TBS, respectively. Based on the AZE, we further propose and demonstrate a new mechanism that induces TBS tunneling by varying measurement strength. Our approach highlights QLM as a versatile tool that can flexibly control wave dynamics. In particular, the AZE can be utilized to mitigate the necessity of adiabaticity, which demands slow evolution or large structure. The AZE is, therefore, a potentially powerful route towards miniaturizing on-chip integrated surface-acoustic-wave and photonic devices. Notably, the AZE may be induced by nonlinear effects [41,42]. The principle shown in our work can be extended to higher-dimensional topological states and be implemented in diverse systems, including both classical wave and single-particle quantum systems.

## Acknowledgment

This work is supported by the National Key R&D Program of China (Grants No. 2023YFA1406900, No. 2022YFA1404403, No. 2021YFB3801800), the National Science Foundation of China (No. 1247043673), the Hong Kong Research Grants Council (RFS2223-2S01, 12301822), Hong Kong Baptist University (RC-RSRG/23-24/SCI/01).

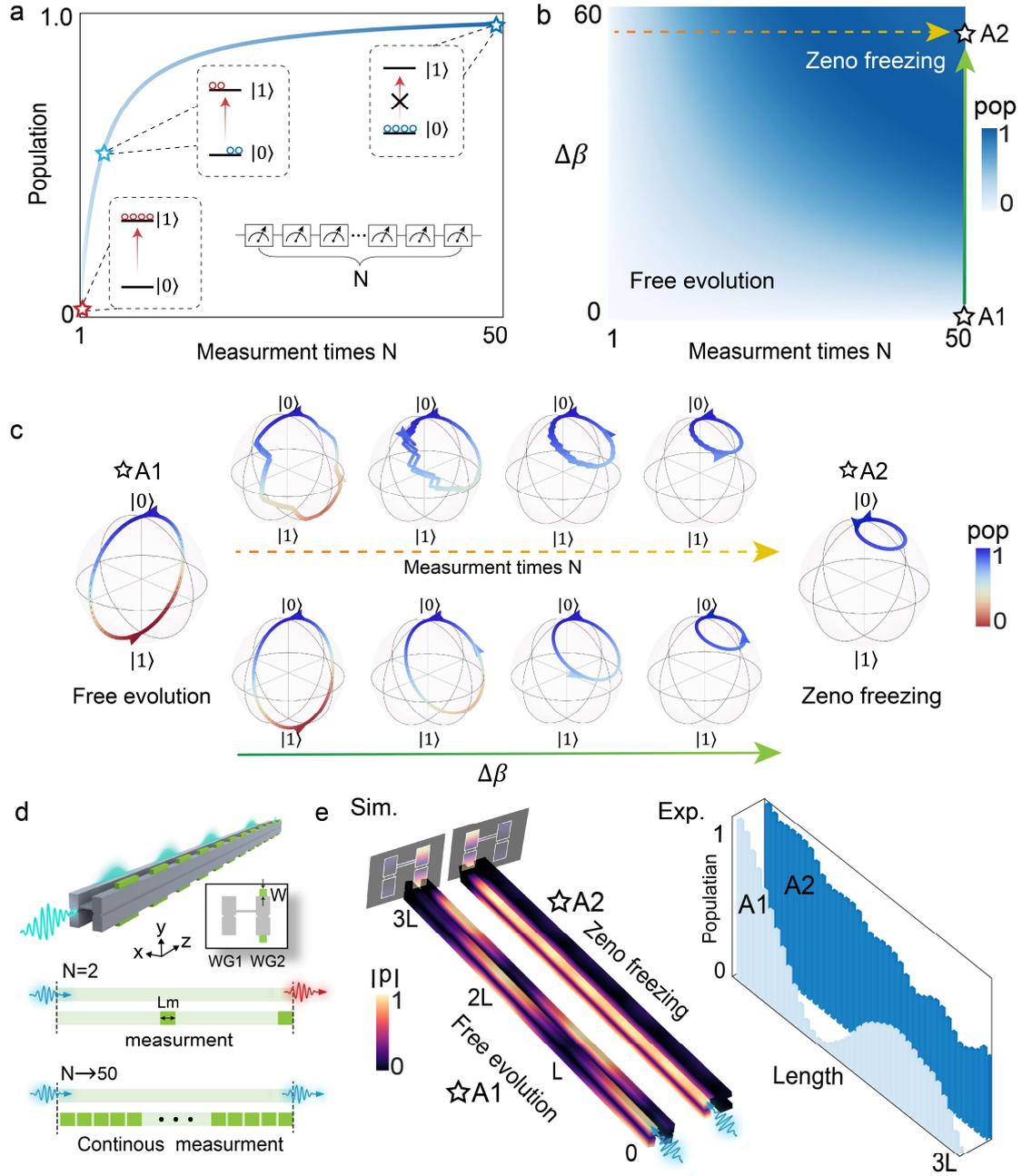

FIG. 1. Zeno effect and its realization in an acoustic two-level system. (a) Repeated QLMs freeze the evolution of a quantum state. (b) The Zeno freezing induced by continuous QLMs in acoustic waveguide systems. The parameters $\beta = 35$ m$^{-1}$, $\kappa = 10.3$, $L = 0.153$ m, typically apply to our system in (d). (c) Evolution trajectories of the initial state on the Bloch sphere along two paths marked in (b). (d) Schematic of the acoustic waveguide system. (e) The acoustic energy population in WG1 and WG2. Simulation and experimental results show typical Rabi oscillation (A1) and the ZE (A2).

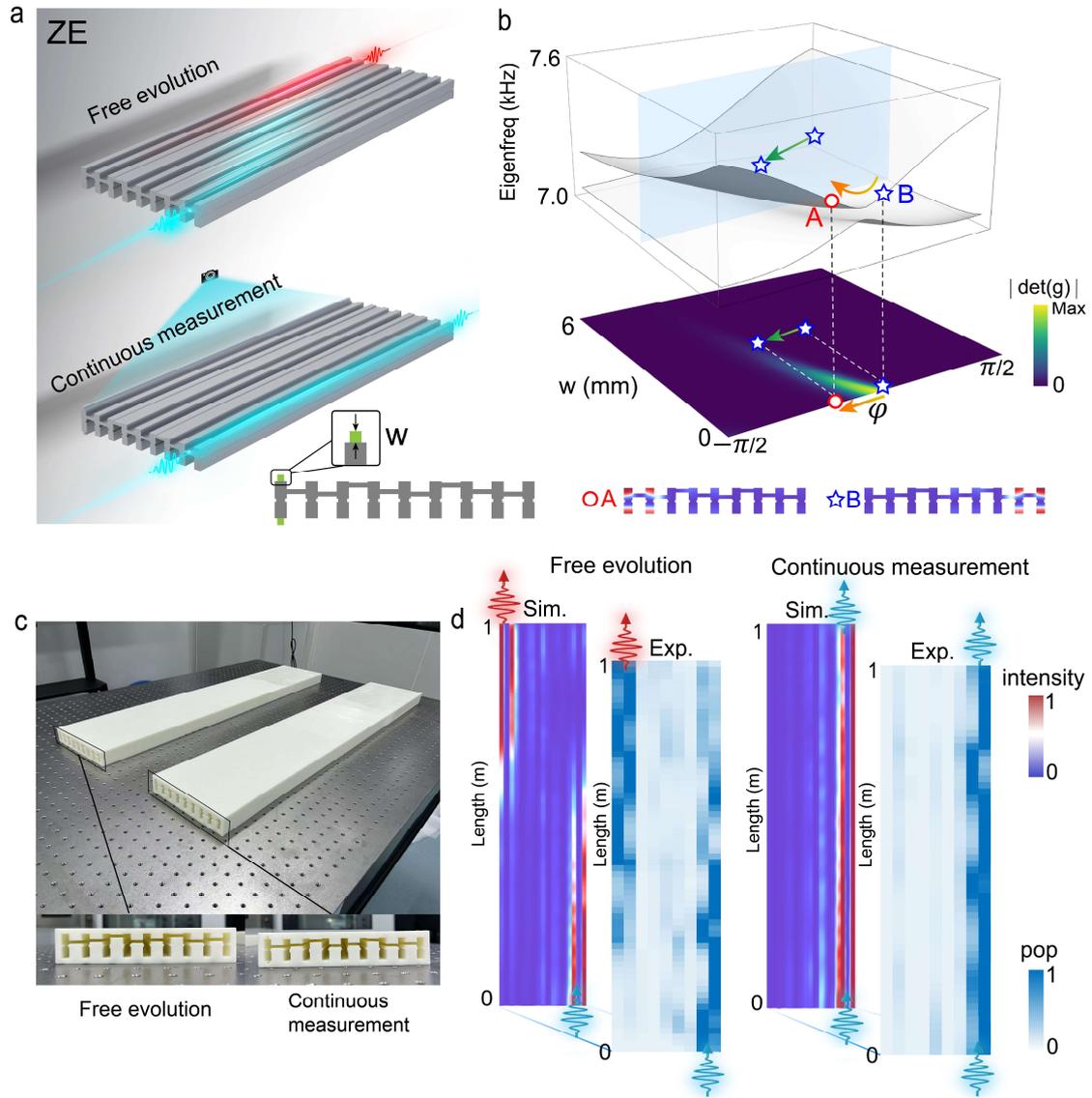

FIG. 2 Zeno freezing of topological boundary states in acoustic system. (a) Illustration of ZE in one-dimensional AAH acoustic waveguide arrays. (b) Impact of continuous QLM on the quantum metric distribution and and its role in eigenstate reconstruction through quantum state distance alterations. The arrows represent the adiabatic evolution paths. (c) Photograph of the experimental acoustic waveguide arrays used to realize the ZE. (d) Simulated and experimental field distributions for free evolution and continuous QLM verifying the ZE.

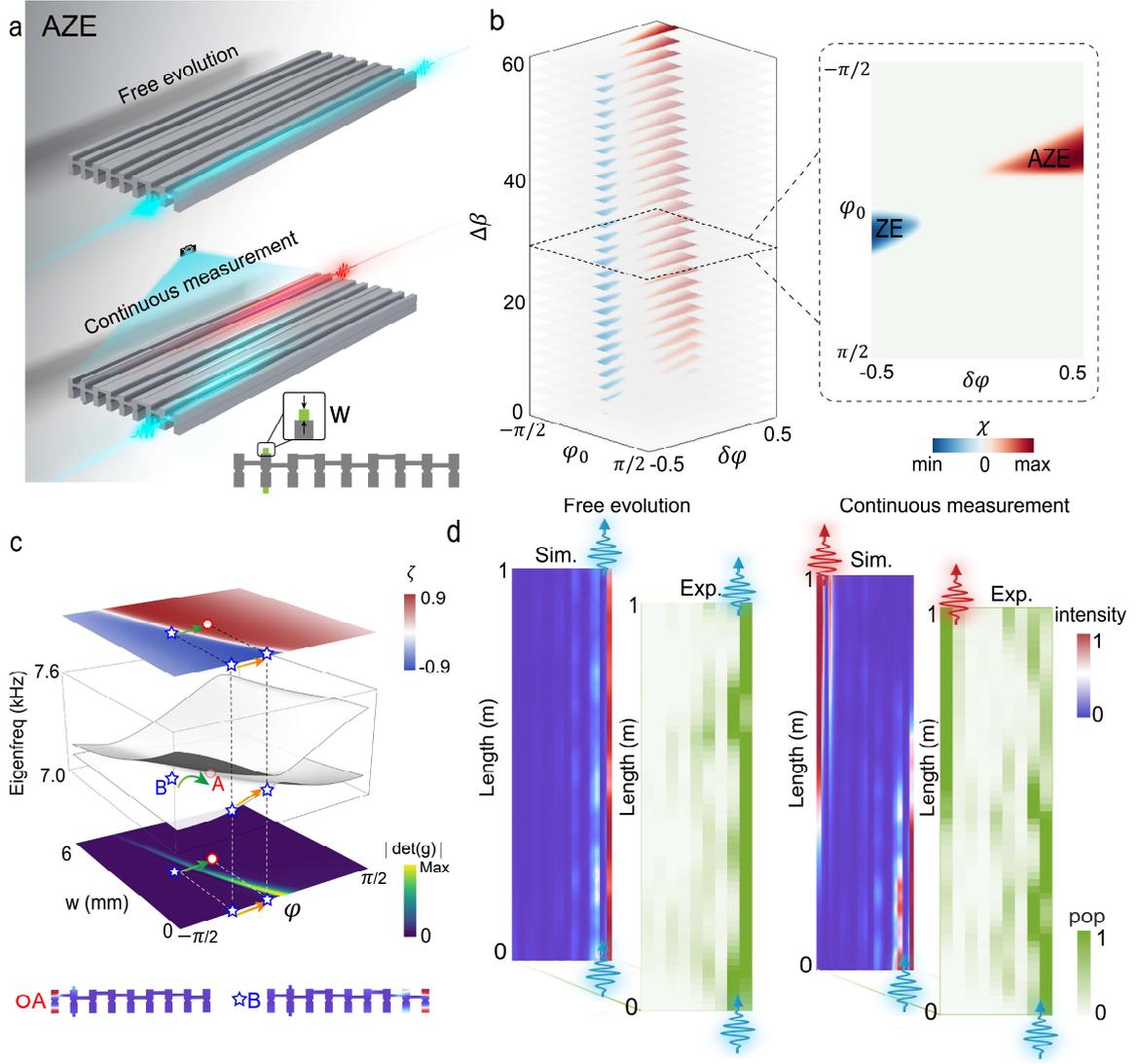

FIG. 3 AZE in topological acoustic waveguide system. (a) Illustration of ZE in one-dimensional AAH acoustic waveguide arrays. (b) The phase diagram shows ZE and AZE regions when QLM is employed on boundary waveguide. The enlarged panel on the right illustrates the distribution of ZE and AZE regions for $w = 3.5$ mm. (c) Impact of QLM on the quantum metric distribution, band structure and eigenvector projection. The arrows represent the adiabatic evolution paths. (d) Simulated and experimental field distributions for free evolution and QLM verifying the AZE.

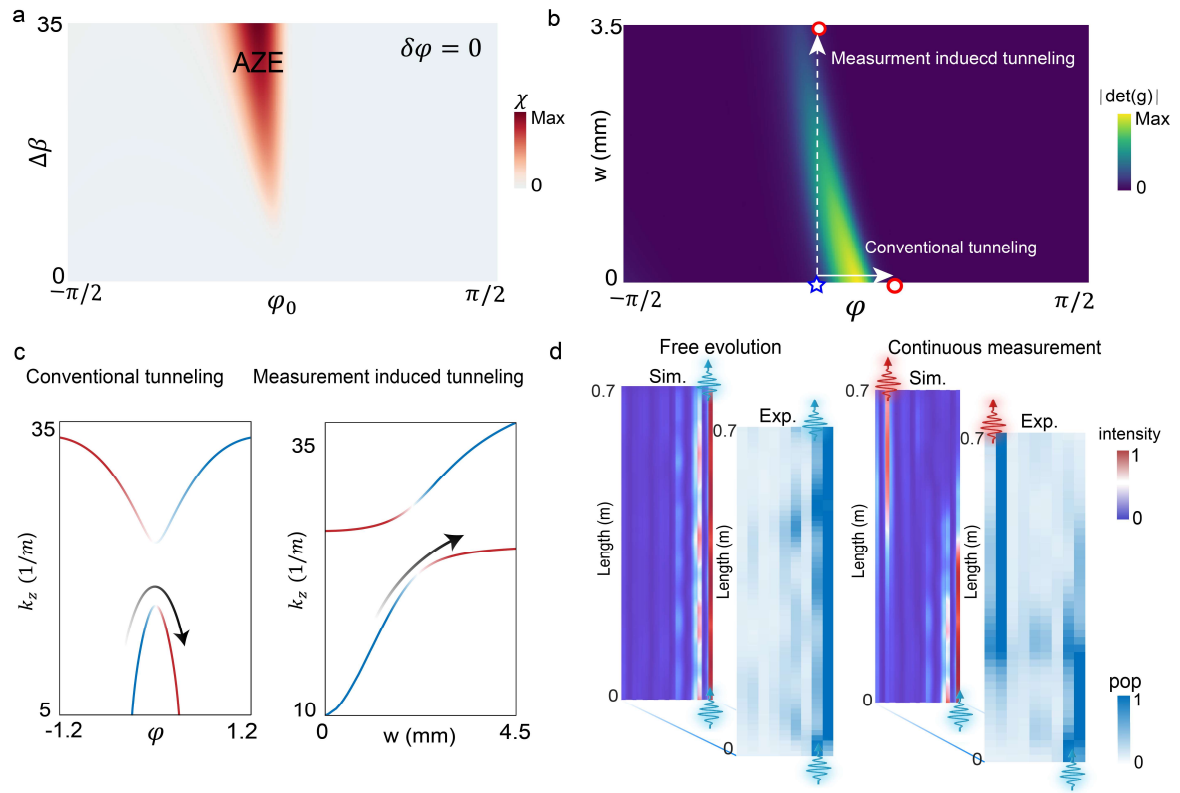

FIG. 4. Topological boundary state transfer induced by varying the measurement strength. (a) Phase diagram implies the accelerated decay of certain initial state through varying the measurement strength. (b) Two distinct tunneling paths marked on the quantum metric $|\det(g)|$ distribution. (c) Comparison of the band structures of these two paths. (d) Simulations and experimental observations of QLM-induced tunneling effect.